\documentclass[useAMS,usenatbib]{mn2e}
\usepackage{graphicx}
\usepackage{subfigure}

\title[Effect of Plasma Composition on the Interpretation of Faraday Rotation ]{Effect of Plasma Composition on the Interpretation of Faraday Rotation }
\author[Kiwan Park and E.G. Blackman]{Kiwan Park$^{1}$\thanks{E-mail: pkiwan@pas.rochester.edu} and E.G. Blackman$^{1}$\thanks{E-mail: blackman@pas.rochester.edu }\\
$^{1}$Department of Physics and Astronomy, University of Rochester, Rochester, NY 14627, USA}
\begin{document}
\date{}
\pagerange{\pageref{firstpage}--\pageref{lastpage}} \pubyear{2009}
\maketitle
\label{firstpage}
\begin{abstract}
Faraday rotation (FR) is widely used  to
 infer the orientation and  strength  of magnetic fields
in astrophysical plasmas.
Although the absence of electron-positron pairs is a plausible assumption in many astrophysical environments,  the  magnetospheres of  pulsars and black holes and their associated jets may involve a significant  pair plasma  fraction. This motivates being mindful of the effect of positrons on FR.  Here we derive and interpret exact expressions of FR for a neutral plasma of arbitrary composition. We focus on electron-ion-positron  plasmas in which  charge neutrality is maintained by an arbitrary combination of ions and positrons.  Because a pure
electron-positron plasma has zero FR, the greater the fraction of positrons the higher the field strength required to account for the same FR.    We  first obtain general formulae and then  specifically  consider parameters relevant to active galctic nuclei (AGN) jets to illustrate the significant differences in  field strengths that FR measurements from radio frequency measurements.
Complementarily, using galaxy cluster core plasmas as examples,  we discuss how  plasma composition can be constrained if independent measurements of the field strength and number density are available and combined with FR.
\end{abstract}
\section{Introduction}

Faraday rotation (FR) describes the effect by which the electric field
vector of propagating electromagnetic (EM) radiation rotates as it propagates
through a magnetized plasma. The  EM waves interact with charged particles
such that the left and right handed components experience different refractive indices
and propagate at different phase velocities. The net electric field vector rotates as the wave propagates.  The total amount of this FR between source
and observer depends on the strength and orientation of the intervening
magnetic field along the line of sight, the plasma density and  the plasma composition.
FR has been used, for example, to  constrain the
the strength, gradient, and direction of magnetic field
 in AGN jets\citep{a1, a2, a4, z1},
 galaxy clusters (REF) and the Galactic ISM  (REF).

Typically, calculations of FR are made for a neutral ion-electron plasma.
While this is a good approximation for  the
plasma of the Galactic ISM and galaxy clusters, it is less reliable
for magnetospheres and outflows around neutron stars and black holes.
In particular, constraining the composition of jets in active galactic nuclei (AGN)
is observationally difficult  because only the electrons or positrons radiate
efficiently so free energy in protons is hard to directly to detect.
The plausible  predominance of electron-plasma in the magnetosphere of the black hole and
the uncertainty as to whether jets emanate directly from its magnetosphere \citep{bz77}
or from the surrounding accretion disk \citep{lop99}
has contributed to making the composition determination a long standing puzzle.
The basic question of how far a jet would be  force-free (\cite{li06}) is also a fundamental issue.
Indirect theoretical constraints on AGN jet plasma composition reach mixed conclusions:
\cite{c1} and \cite{g08}  favor  predominately electron-proton jets,
whilst \cite{r1} favors a predominantly electron-positron jet for
M87. \cite{hubbard} argued that on the largest scales, stellar wind mass loading
will significantly proton load the jet regardless of its initial composition.
It is therefore important to identify possible new techniques for determining jet plasma composition and to be aware of its effect on measurements such as FR.

With the above motivations in mind,  we calculate the effect of plasma composition on FR from first principles.
In section 2 we first derive the general exact formula for FR in an arbitrary
neutral plasmas. We then consider the special cases of a pure ion electron plasmas
and an ion-electron-positron plasma.  In section 3 we solve the exact and high frequency
approximations to the general FR equation and show quantitatively how plasma composition
leads to degeneracies in the electron number density $n_e$, the magnetic field $B$.  We
discuss how these degeneracies can in principle be used to constrain the plasma
composition in AGN jets,  and conclude in section 4.

\section{ Generalized Faraday rotation }
\subsection{Formalism for Arbitrary Neutral Plasmas}
To formally derive FR for an arbitrary neutral plasma, we assume a cold neutral plasma
in a background external magnetic field $\vec{B}_{ex}$, subject to a  perturbation from a propagating electromagnetic wave.  (The cold plasma approximation for
FR has been shown to be effective for the electron contribution to
FR even for quasi-relativistic plasmas \citep{s71}, and we discuss this further below Eq.(\ref{16aa})).
For the
electric field $\vec{E}$,  magnetic field  $\vec{B}$ and induced
particle velocity $\vec{v}_s$  (where the index $s$ indicates particle species) we write
\begin{eqnarray}
\label{FR derivation assumption}
&&\vec{E}=\vec{E}_{1},\nonumber\\
&&\vec{B}=\vec{B}_{1}+\vec{B}_{ex},\nonumber\\
&&\vec{v}_{s}=
\vec{v}_{s1},
\end{eqnarray}
where $\vec{E}_1$ and $\vec{B}_1$ are perturbations such that
$|{B}_{1}/B_{ex}|\ll1$,  $|{E}_{1}/B_{ex}|\ll1$ and ${v}_{s1}|\ll c$.
We also assume a neutral plasma so that
\begin{equation}
\quad\sum_{s} n_{s0}e_{s}=0
\label{neutral},
\end{equation}
where $n_{s0}$ is the unperturbed density of a particle of species $s$, and $e_{s}$ is charge of particle of species $s$.

Using the above formalism, Maxwell equations become (e.g. \cite{g1}):
\begin{eqnarray}\label{FR derivation1}
\nabla \times \vec{E_{1}}=-\frac{1}{c}\frac{\partial \vec{B_{1}}}{\partial t}, \quad
\nabla \times \vec{B_{1}}=\frac{4\pi}{c}\vec{J}+\frac{1}{c}
\frac{\partial \vec{E_{1}}}{\partial t}.
\end{eqnarray}
If the current density $\vec{J}\equiv \sum_s n_{s0}e_{s}\vec{v}_{s1}=0$,  then $E_{1}$ \& $B_{1}$ are decoupled, resulting in the plane wave vacuum equations.  However
a finite $\vec{J}$  and the Lorentz force equation
\begin{equation}
m_{s} \frac{d\vec{v}_{s}}{dt}=e_{s}(\vec{E}_{1}+\frac{1}{c}
\vec{v}_{s}\times \vec{B}_{ex})
\label{4}
\end{equation}
imply that in general, that $\vec{E}$, $ \vec{B}$, and $\vec{v}_s$ are all coupled.

To quantify the interaction between the particles and EM fields,
we take $\vec{v}_s, \vec{E}_1 \propto e^{i(\vec{k}\cdot \vec{x}-\omega t)}$
so  Eq.(\ref{4}) becomes
\begin{eqnarray}\label{Lorentz force}
\vec{v}_{s}=\frac{i e_{s}}{m_{s}\omega}(\vec{E}_1+\frac{\vec{v}_{s}}{c}\times
\vec{B}_{ex})=\frac{i e_{s}}{m_{s}\omega}(\vec{E}_1+\frac{\vec{v}_{s}}{c}\times
\frac{m_{s}c\,\vec{\omega}_{cs}}{e_{s}}),
\end{eqnarray}
where $\omega_{cs}=e_{s}B_{ex}/m_{s}c$ is the cyclotron frequency of species $s$.
The current density
can then be expressed as the product of a conductivity tensor and the  $\vec{E}_1$ field, namely
$\vec{J}=\overleftarrow{\overrightarrow{\sigma}} \cdot \vec{E_{1}}$, where
 the  conductivity tensor is given by \citep{a3}
\begin{eqnarray}\label{conductance}
\sigma_{ij}=\sum_{s}\frac{i\,n_{s0}e_{s}^2}{m_{s}
\omega[1-(\frac{\omega_{cs}}{\omega})^{2}]}
\left(\delta_{ij}-\frac{\omega_{cs, i} \omega_{cs, j}}{\omega^{2}}-
\frac{i}{\omega}\epsilon_{ijk}\omega_{cs, k}\right).&\nonumber\\
\end{eqnarray}

We now take $\vec{B}_{ex}=(0,0,B)$
so that the components of  Eq.(\ref{Lorentz force}) become
\begin{eqnarray} \label{Lorentz force 2}
&&-i\omega m_{s} v_{sx}=e_{s}(E_{1x}+\frac{v_{sy}}{c} B),\nonumber\\
&&-i\omega m_{s} v_{sy}=e_{s}(E_{1y}-\frac{v_{sx}}{c} B),\nonumber\\
&&-i\omega m_{s} v_{sz}=e_{s}E_{1z},
\end{eqnarray}
and, the conductivity tensor becomes
\begin{eqnarray}\label{conductance-Matrix}
\overleftarrow{\overrightarrow{\sigma}}=\sum_{s}\frac{n_{s0}e_{s}^{2}}{m_{s}}
\left(\begin{array}{ccc}
\frac{-i \omega}{\omega_{cs}^{2}-\omega^{2}} &
\frac{\omega_{cs}}{\omega_{cs}^{2}-\omega^{2}}  &
0 \\ \frac{-\omega_{cs}}{\omega_{cs}^{2}-\omega^{2}}  &
\frac{-i \omega}{\omega_{cs}^{2}-\omega^{2}} & 0 \\
0 & 0 & \frac{i}{\omega}\end{array}\right).
\end{eqnarray}
 Eq.(\ref{FR derivation1}) with $\vec{\nabla} \rightarrow i \vec{k}$ then becomes
\begin{eqnarray}\label{FR dispersion relation-Vector}
\vec{k} \times (\vec{k} \times \vec{E}_{1})+
\frac{\omega^{2}}{c^{2}}(1-\frac{4\pi \overleftarrow{\overrightarrow{\sigma}}}
{i\omega})\cdot\vec{E}_{1}=0.
\end{eqnarray}
from which the secular equation for the FR effect is
\begin{eqnarray}\label{FR dispersion relation-Matrix}
\left(\begin{array}{ccc}\label{secular equation of E field}
S-n^{2} &
-iD & 0 \\
iD & S-n^{2} & 0 \\
0 & 0 & P
\end{array}\right)
\left(
  \begin{array}{c}
    E_{1x} \\
    E_{1y} \\
    E_{1z}
  \end{array}
\right)=0,
  \end{eqnarray}
where  $S\equiv1-\sum_{s}\frac{\omega_{ps}^{2}}{\omega^{2}-\omega_{cs}^{2}}$,
$D\equiv\sum_{s}\frac{\omega_{cs}\omega_{ps}^{2}}{\omega(\omega^{2}-\omega_{cs}^{2})}$,
$P\equiv1-\sum_{s}\frac{\omega_{ps}^{2}}{\omega^{2}}$ and the plasma
frequency and wave vector are given respectively by
$ \omega_{ps}^{2}=\frac{4\pi n_{s}e_{s}^{2}}{m_{s}}$ and
$ \vec{k} = \frac{n\omega}{c} \, \hat{z}$.

The solution of  Eq.(\ref{FR dispersion relation-Matrix})
for refractive index $n$ produces non-trivial FR when the
the two solutions for $n$ are distinct, corresponding to left and right handed
polarizations;  $n^{2}_{L}$=$S-D$, $n^{2}_{R}$=$S+D$
with two associated $\vec{E}_1$ field eigenvectors.
The transverse $(x,y)$  components of $\vec{E}$ for each refractive index
are of the same magnitude but have different phases, that is,
$E_{L}=(E_{0}, -iE_{0}, 0)$, $E_{R}=(E_{0}, iE_{0}, 0)$, where $\pm i$ arises from the differentiation of velocity and position over time in the Lorentz force law. The equal
amplitude of the transverse $E$ field components then imply
 circularly polarized waves.

The different phase velocities ($c/n_{L},\,c/n_{R}$) cause
the propagating left and right handed circularly polarized waves to experience
a net phase angle difference when they propagate over the same distance.
 As a result, the net electric field  phase angle ($\phi=Tan^{-1}(E_{y}/E_{x})$)
 that comes from the superposition of these handed  waves
 rotates  along the propagating distance.  This is the  FR.
 The change $\phi$ along the propagation distance is
\begin{eqnarray}\label{Theoretical FR expression1}
\frac{d\phi}{dz}&=&\frac{1}{2}(k_{L}-k_{R})\nonumber\\
&=&\frac{\omega}{2c}\Bigg(\sqrt{1-\sum_{s}\frac{\omega_{ps}^{2}}
{\omega (\omega-\omega_{cs})}}
-\sqrt{1-\sum_{s}\frac{\omega_{ps}^{2}}{\omega (\omega+\omega_{cs})}}\,\Bigg).\nonumber\\
\end{eqnarray}

Having derived the general formalism for a neutral plasma of arbitrary composition
 and the exact equation for  FR (Eq.(\ref{Theoretical FR expression1})),  we note that
\cite{h1} considered FR in an ion-electron positron plasma
producing  the approximate analytical result
\begin{eqnarray} \label{FR approximate equation Shukla}
\frac{d\phi}{dz}\sim Z_{i}n_{i}
\frac{2\pi e^{3}B}{m_{e}^{2}c^{2}\omega^{2}},
\end{eqnarray}
where $Z_i$ is the ion charge number.
Eq.(\ref{FR approximate equation Shukla}) agrees with Eq.(\ref{Theoretical FR expression1}) in the high frequency limit for an ion-electron-positron plasma, a point we will return to in section 2.3.
Eq.(\ref{FR approximate equation Shukla}) indicates there is no FR
in case of an electron-positron pair plasma($n_{i}=0$).
For $n_{i}\neq 0$, ions generate the FR by breaking the symmetry of a pair plasma.

\subsection{Ion-electron plasma}
For a pure (hydrogen) ion-electron plasma,
Eq.(\ref{neutral}) takes the form
$n_e +n_i =0$,  where $n_e$ and $n_i$ are the electron and hydrogen ion number densities.
The summation over $s$ in Eq.(\ref{Theoretical FR expression1})
also involves terms corresponding to electrons and ions respectively.
However, because of the  large ion to electron mass ratio $m_i/m_e =1836$,
the ion terms are typically ignored  (e.g.\cite{a1, a2, z1}).
Then the  rotated angle
integrated along the line of sight for the distance $l$ (e.g. \cite{r1}) becomes
\begin{eqnarray}\label{1st order expanded FR}
\phi &\simeq& \int_{0}^{l}\frac{\omega}{2c}\Big(\sqrt{1-\frac{\omega_{pe}^{2}}{\omega
(\omega-\omega_{ce})}}
-\sqrt{1-\frac{\omega_{pe}^{2}}{\omega (\omega+\omega_{ce})}}\,\,\Big)\cdot dz \nonumber\\
&\sim& \frac{2\pi e^{3}}{m_{e}^{2}c^{2}\omega^{2}}
\int_{0}^{l} n_{e}B\cos\theta \: dz\nonumber\\
&=&\Big(\frac{e^{3}}{2\pi m_{e}^{2}c^{4}}\int_{0}^{l} n_{e}B\cos\theta \: dz\Big)
\cdot \lambda^{2}\equiv RM \cdot \lambda^{2},
\end{eqnarray}

where the second relation follows for $\omega \gg \omega_{pe},\omega_{ce}$.
The general procedure for determining RM is to measure $\phi$ at multiple wavelengths
and infer a slope of the  $\phi$ vs. $\lambda$  line.

\subsection{Ion-electron-positron plasma}

\noindent
For the hydrogen ion-electron-positron case Eq.(\ref{neutral}) becomes
\begin{eqnarray}
-en_{e}+en_{e^{+}}+en_{i}=0,
\end{eqnarray}
where $n_{e^+}$ is the positron number density.
We now define $X \equiv n_{e^{+}}/n_{e}$ so that
\begin{eqnarray}\label{definition of densities}
n_{e^{+}}=n_{e}X,\quad n_{i}=n_{e}(1-n_{e^{+}}/n_{e})=n_{e}(1-X).
\end{eqnarray}
Eq.(\ref{Theoretical FR expression1}) then becomes
\begin{eqnarray}\label{exact FR expression}
&&\frac{d\phi}{dz}=\frac{\omega}{2c}\Bigg(\sqrt{ 1-\frac{\omega_{pe}^{2}}{\omega (\omega-\omega_{ce})}-\frac{\omega_{pe^{+}}^{2}}{\omega (\omega-\omega_{ce^{+}})}-\frac{\omega_{pi}^{2}}{\omega (\omega-\omega_{ci})} }\nonumber\\
&&-\sqrt{ 1-\frac{\omega_{pe}^{2}}{\omega (\omega+\omega_{ce})}-\frac{\omega_{pe^{+}}^{2}}{\omega (\omega+\omega_{ce^{+}})}-\frac{\omega_{pi}^{2}}{\omega (\omega+\omega_{ci})} }\;\Bigg)\nonumber\\
&&=\frac{\omega}{2c}\Bigg(\sqrt{1- \frac{4\pi e^{2}}{m_{e}}\frac{n_{e}}{\omega}  \Big( \frac{1}{\omega+\frac{eB}{m_{e}c}}+\frac{X}{\omega-\frac{eB}{m_{e}c}}
+\frac{1-X}{1836\omega-\frac{eB}{m_{e}c}}  \Big) }\nonumber\\
&&-\sqrt{1- \frac{4\pi e^{2}}{m_{e}}\frac{n_{e}}{\omega}  \Big( \frac{1}{\omega-\frac{eB}{m_{e}c}}+\frac{X}{\omega+\frac{eB}{m_{e}c}}
+\frac{1-X}{1836\omega+\frac{eB}{m_{e}c}}  \Big)}  \Bigg)\nonumber\\
&&=\frac{\omega}{2c}\Big(\sqrt{1-q_{L}}-\sqrt{1-q_{R}}\Big),
\label{16aa}
\end{eqnarray}
where $q_L$ represents the second term under the first square root of the second equality and $q_R$ represents the second term under the second square root of
the second equality.

Eq.(\ref{Theoretical FR expression1}), (\ref{FR approximate equation Shukla}) and (\ref{exact FR expression}) presume a cold plasma and it is instructive to comment on the validity
of these expressions for a  warm plasma. For the latter,
 motions of charged particles are influenced
by thermal effects in addition to the electromagnetic force
and to express the current density the solution of Vlasov equation is necessary
(e.g. \cite{g1}, chapt. 9).
\cite{s71} studied FR for a warm ion-electron plasma and found that for large frequencies away from resonances, the  correction to the electron contribution to FR  is small compared to the cold plasma term. However, for a pair plasma in which the electron and positron cold plasma terms cancel exactly, warm plasma correction terms would not cancel exactly and a finite contribution
would remain as the positron and electron correction terms do not cancel. We ignore these small  corrections for present purposes and leave further discussion for future work.

In the high frequency limit, $q_{L}\ll1$ and $q_{R}\ll1$,
and we can approximate Eq.(\ref{exact FR expression}) as
\begin{eqnarray} \label{FR approximate equation}
\frac{d\phi}{dz}&\simeq& \frac{2\pi e^{3}}{m_{e}^{2}c^{2}}n_{e}B(1-X)
\bigg( \frac{1}{\omega^{2}-(\frac{eB}{m_{e}c})^{2} }-
\frac{1}{1836^{2} \omega^{2}-(\frac{eB}{m_{e}c})^{2} }  \bigg)\nonumber\\
&\sim& \frac{2\pi e^{3}}{m_{e}^{2}c^{2}}n_{e}B(1-X)
(\frac{1}{\omega^{2}}-\frac{1}{1836^{2}\omega^{2}}).
\end{eqnarray}

Using Eq.(\ref{definition of densities}),
then Eq.(\ref{FR approximate equation}) is the same as
Eq.(\ref{FR approximate equation Shukla}).

For a pure neutral pair plasma ($X$=1),
the right side of Eq.(\ref{exact FR expression})  (or (\ref{FR approximate equation}))
 vanishes. The  FR vanishes because the equal mass of positrons and electrons
induce the same phase speeds for oppositely handed EM waves.
This contrasts the limit  of the previous subsection of a pure neutral  (hydrogen) ion-electron plasma ($X=0$), for which the mass asymmetry leads to unequal phase speeds of the oppositely handed waves and a finite right side of Eq.(\ref{exact FR expression}).  In general,  for $0\le X \le 1$  with  $n_e$ , $B_{ex}$ and source distance fixed,
the right side of Eq.(\ref{exact FR expression})  decreases with increasing $X$.
We discuss  solutions of Eq.(\ref{exact FR expression}) in the next section.

Note that the exact FR expression (\ref {exact FR expression})  has singularities when the wave frequency
of the EM wave coincides with the particle cyclotron frequencies, i.e.
 at $\omega=eB/m_{e}c=1.76\times10^{7}\,B$ (for electrons and positrons) and $eB/m_{i}c=9571\,B$ (for ions).   The FR  would exhibit sharp resonance features
near the singular points, allowing the
 $B$ field to be inferred in principle.  However, for applications to
 extended jets of AGN and larger scale systems, these resonant frequencies
 are generally small compared to the relevant $\sim$ GHz  frequencies.

\section{Astrophysical Implications}
\begin{figure*}
\centering
\mbox{
   \subfigure[log $B$ vs $n_{e}$]
   {
     \includegraphics[width=6cm]{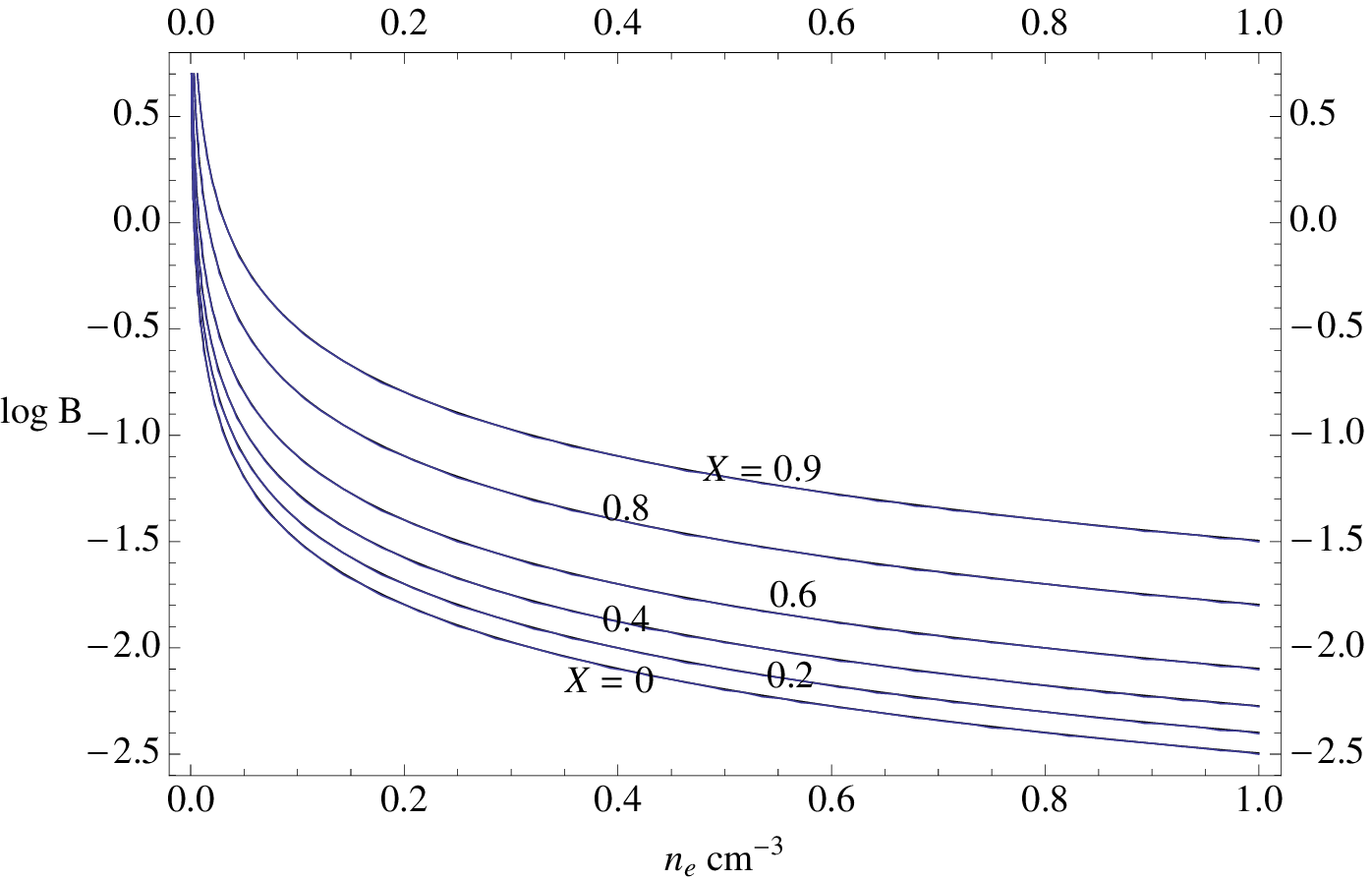}
     \label{fig:N vs ln B}
   }
   \subfigure[log $B$ vs $X$]
   {
     \includegraphics[width=6cm]{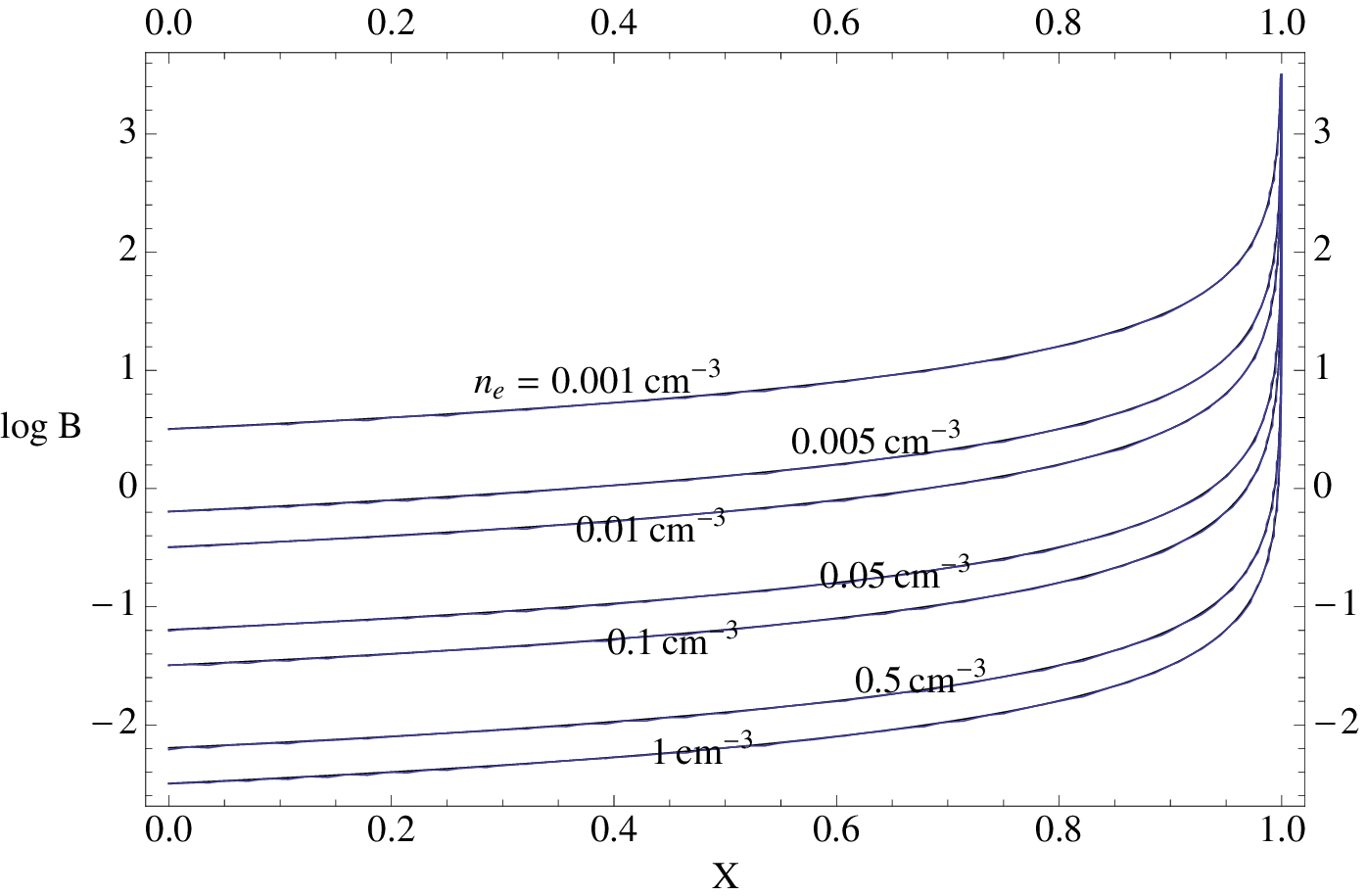}
     \label{fig:X vs ln B}
   }
   \subfigure[log $n_{e}$ vs $X$]
   {
     \includegraphics[width=6cm]{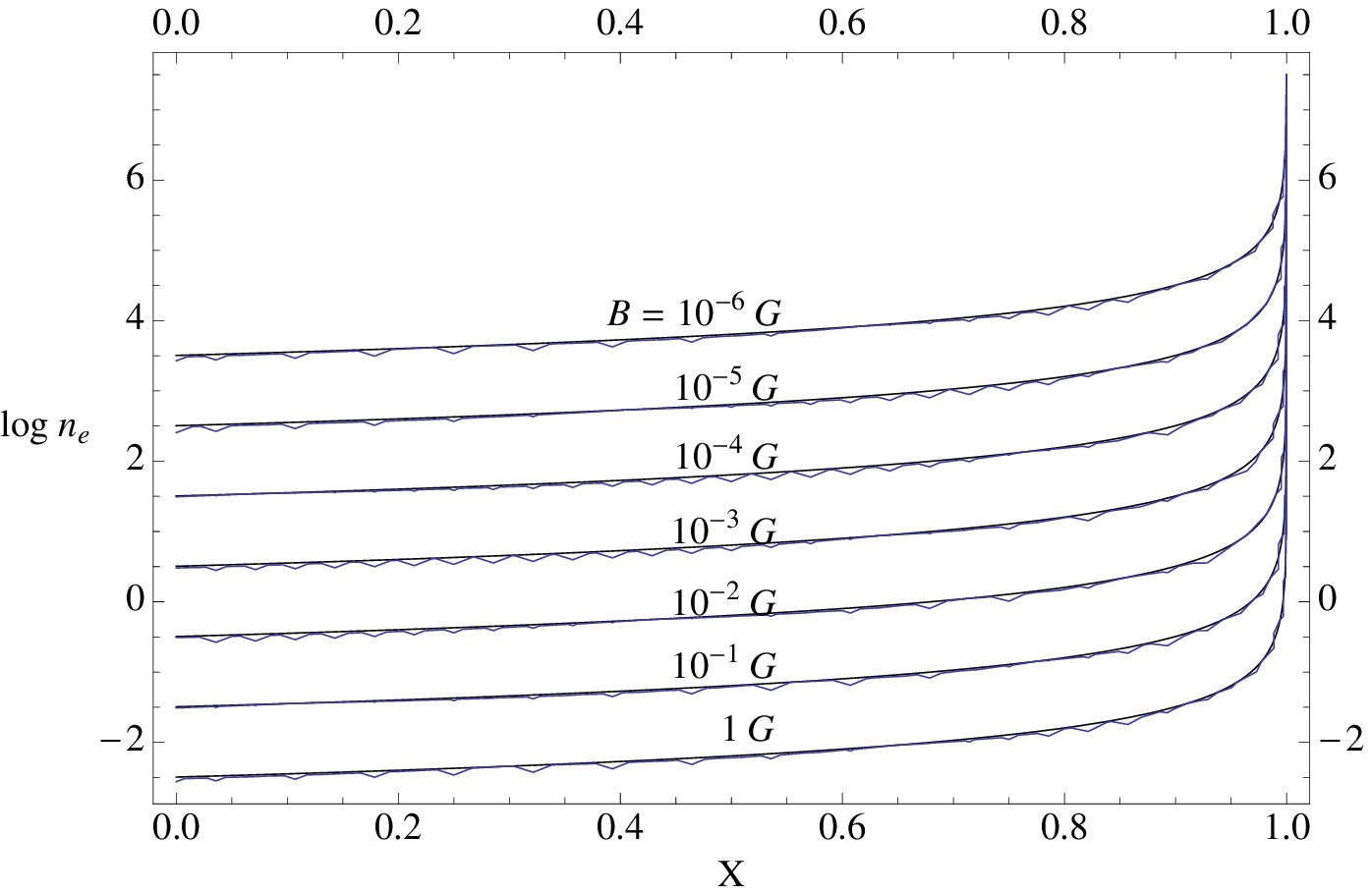}
     \label{fig:X vs ln N}
   }

    }

\caption{Plots of $n_{e}$, $B$ and $X$. All panels use a fixed  RM = 2592 rad $m^{-2}$ in Zavala \& Taylor (2001) (Fig.3). The line of sight $B$ field is estimated at $\sim 0.5$ mas ($\sim 3.63$ pc) from the core.
All of the plots were  made using both Eq.(\ref{exact FR expression})
and Eq.(\ref{FR approximate equation}) with $\lambda$=1.35 $cm$ (22.2 GHz).
The plots of each equation were overlapped highlighting the efficacy of the approximate
equation. The original observation frequency is 8GHz but any  frequency  gives the same RM.
}
\end{figure*}

\begin{figure*}
\centering
\mbox{
     \subfigure[log $B$ vs $n_{e}$]
   {
     \includegraphics[width=6cm]{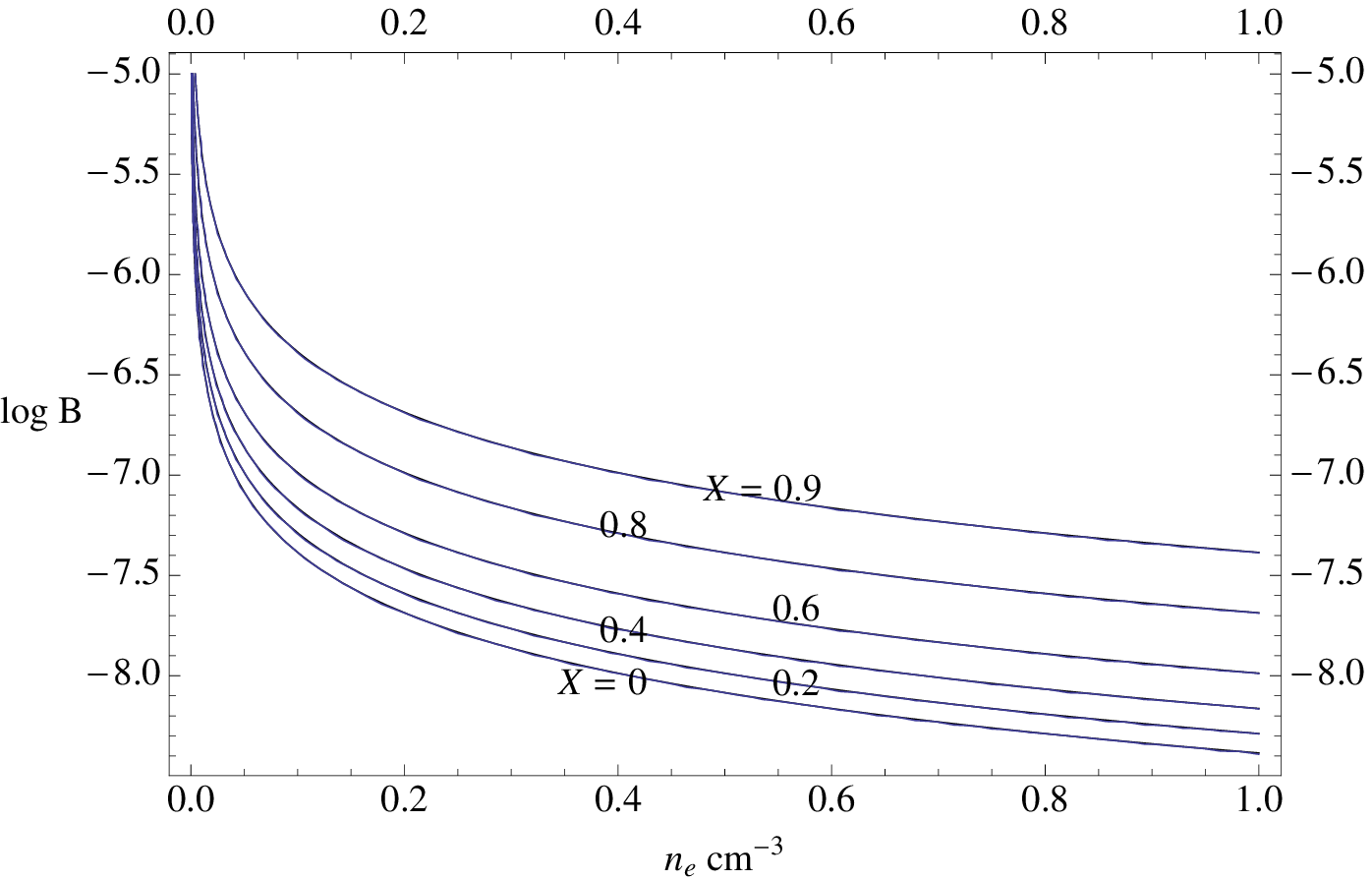}
     \label{fig:N vs ln B cluster}
   }
   \subfigure[log $B$ vs $X$]
   {
     \includegraphics[width=6cm]{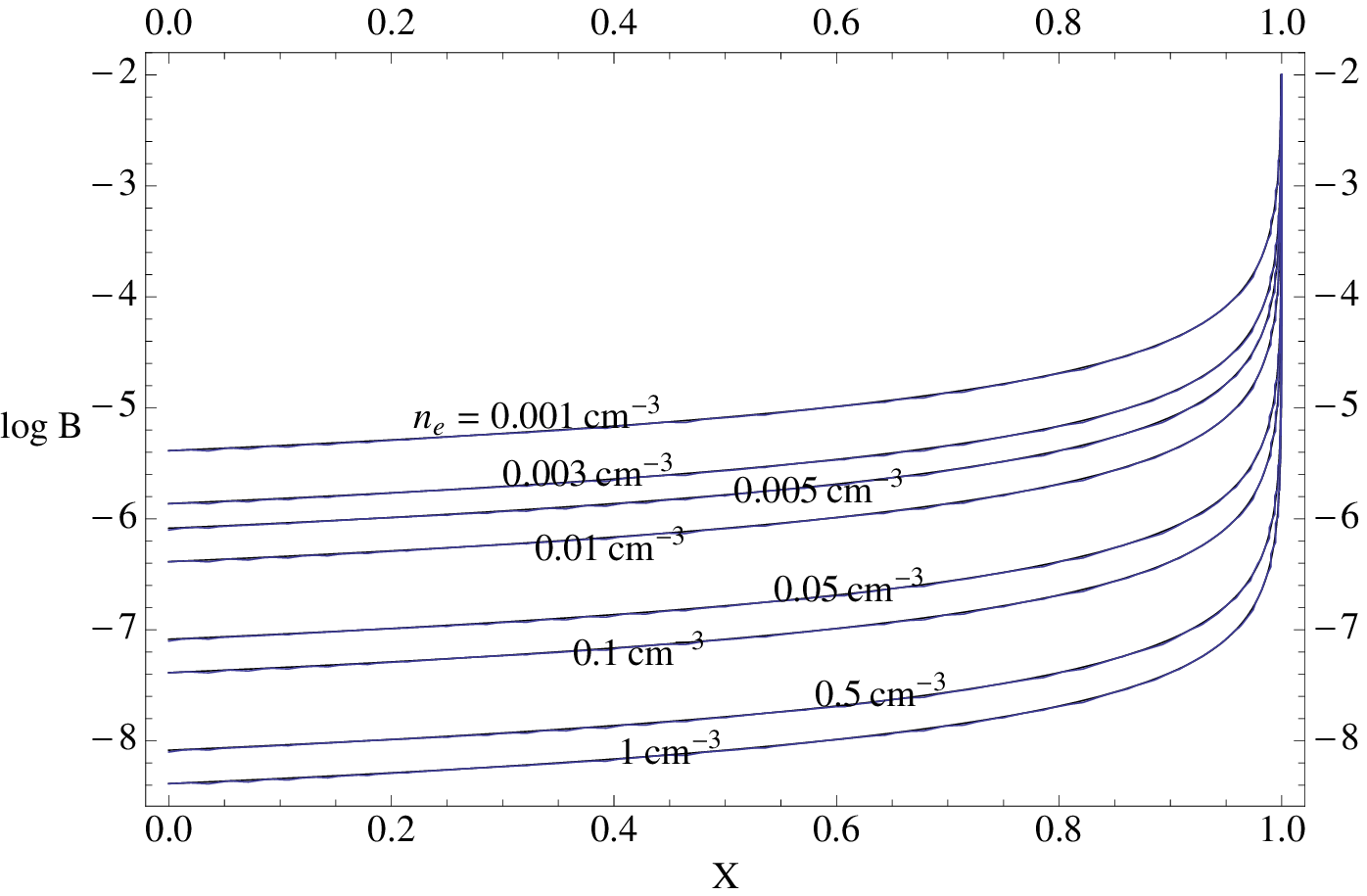}
     \label{fig:X vs ln B cluster}
   }
   \subfigure[log $n_{e}$ vs $X$]
   {
     \includegraphics[width=6cm]{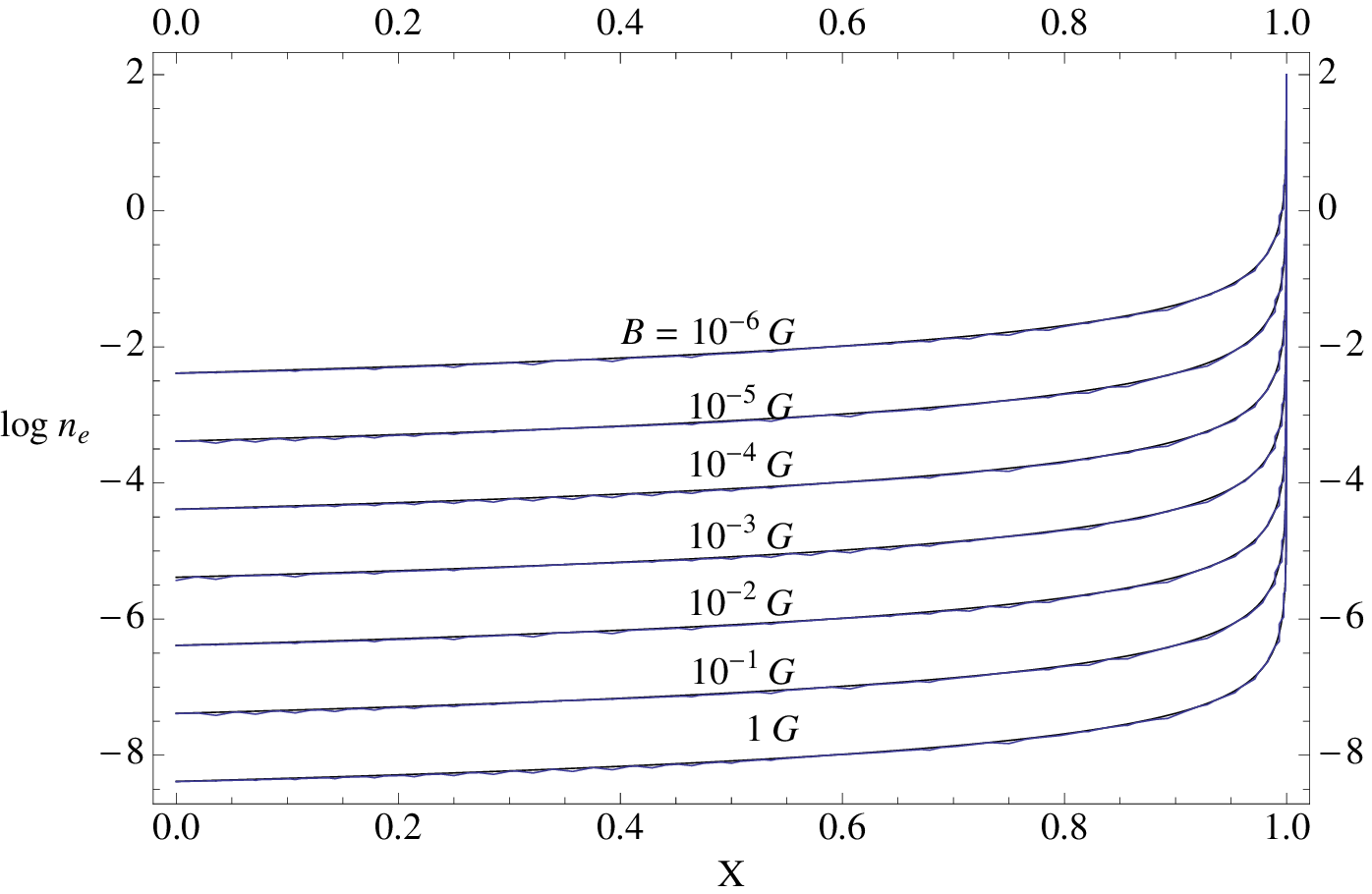}
     \label{fig:X vs ln N cluster}
   }

    }

\caption{Same as Fig.1. but using RM$\sim$1000 rad $m^{-2}$, path length $l = $300 kpc,
in an X-ray cluster core Jaffe (1980) with $\lambda$=1.35 $cm$ (22.2 GHz)
Again both Eq.(\ref{exact FR expression}) and Eq.(\ref{FR approximate equation})
were used for the plots.
}
\end{figure*}

\subsection{General  Implications of Plasma Composition}

Figs.1 and 2 show  solutions
to the exact expression  Eq.(\ref{exact FR expression})
and the approximation  Eq.(\ref{FR approximate equation}) for $n_{e}$, $B$ and $X$
at fixed values  of  RMs (2592 rad/m$^{2}$ in Fig.1. and 1000 rad/m$^{2}$ in Fig.2).
The RMs were converted into rotation angles for $cm$ scale
at $\lambda$= 1.35 $cm$(22.2GHz).
The FR in Fig.1
corresponds to 3.63 pc (749 Mpc $\times$ 0.5 mas, C1 region) from the core of AGN jet 3C 273 from \cite{z2}.
From synchrotron emission, \cite{t1} calculated the total magnetic field
 to be ($B\sim0.06$ G)  in this region.  If this were the line of sight field,
Fig.\ref{fig:X vs ln B} shows  that  $n_{e}\geq$0.05 $cm^{-3}$  for RM (2592 rad m$^{-2}$),
the minimum $n_e$ occurring at $X=0$.
Fig.\ref{fig:X vs ln N cluster}
shows  a complementary  example for values appropriate for a  typical X-ray cluster (\cite{j1})
with
$n_{e}\sim$ 0.003 $cm^{-3}$ and RM $\sim$ 1000 rad $m^{-2}$.
 The standard assumption that $X=0$ for a known distance leads directly to the inference that
$B\sim1 \mu$G\citep{j1}.  But for  any $X<1$,
Fig.\ref{fig:X vs ln B cluster} for example,  shows how much
stronger the  field could be.

For fixed values of RM, Figs.\ref{fig:N vs ln B}, and \ref{fig:N vs ln B cluster} show that $n_{e}$
and $B$ behave oppositely for  each value of  $X$: As $n_{e}$ increases (decreases),
$B$ decreases (increases). These trends reflect  that the RM is roughly proportional to the
field and the density.   Figs.\ref{fig:N vs ln B}, \ref{fig:X vs ln B}, \ref{fig:N vs ln B cluster},
and \ref{fig:X vs ln B cluster} show that  as $X$ increases (decreases), $B$ and $n_{e}$ increase (decrease) respectively. These trends result because an increasing $X$ means a higher fraction of pair plasma. The latter  contributes zero FR so that a higher $B$ or $n_e$ is needed for a fixed RM.
  Figs.\ref{fig:X vs ln N}, and \ref{fig:X vs ln N cluster} reflect these same trends.

In all panels, the  lines resulting from Eq.(\ref{exact FR expression}) are indistinguishable from those obtained using
Eq.(\ref{FR approximate equation})  for the parameters used, highlighting the efficacy
of the latter.
Figs.\ref{fig:N vs ln B cluster},
\ref{fig:X vs ln B cluster} and \ref{fig:X vs ln N cluster} are very similar to Figs.\ref{fig:N vs ln B}, \ref{fig:X vs ln B} and \ref{fig:X vs ln N} as
only the vertical axis scales are different due to the different  RM choices.
Overall, the figures show  how an unknown plasma composition $X$ implies degeneracies
in the values  $n_e$ and $B$, or complementarily, how independent measures of
$n_e$ and $B$ can be combined with an RM to constrain $X$.

\subsection{Further Discussion of Applications}
\begin{figure*}
\centering
\mbox{
   \subfigure[log $B$ vs $X$ for M87 (Zavala \& Taylor 2002)]
   {
      \includegraphics[width=8.5cm]{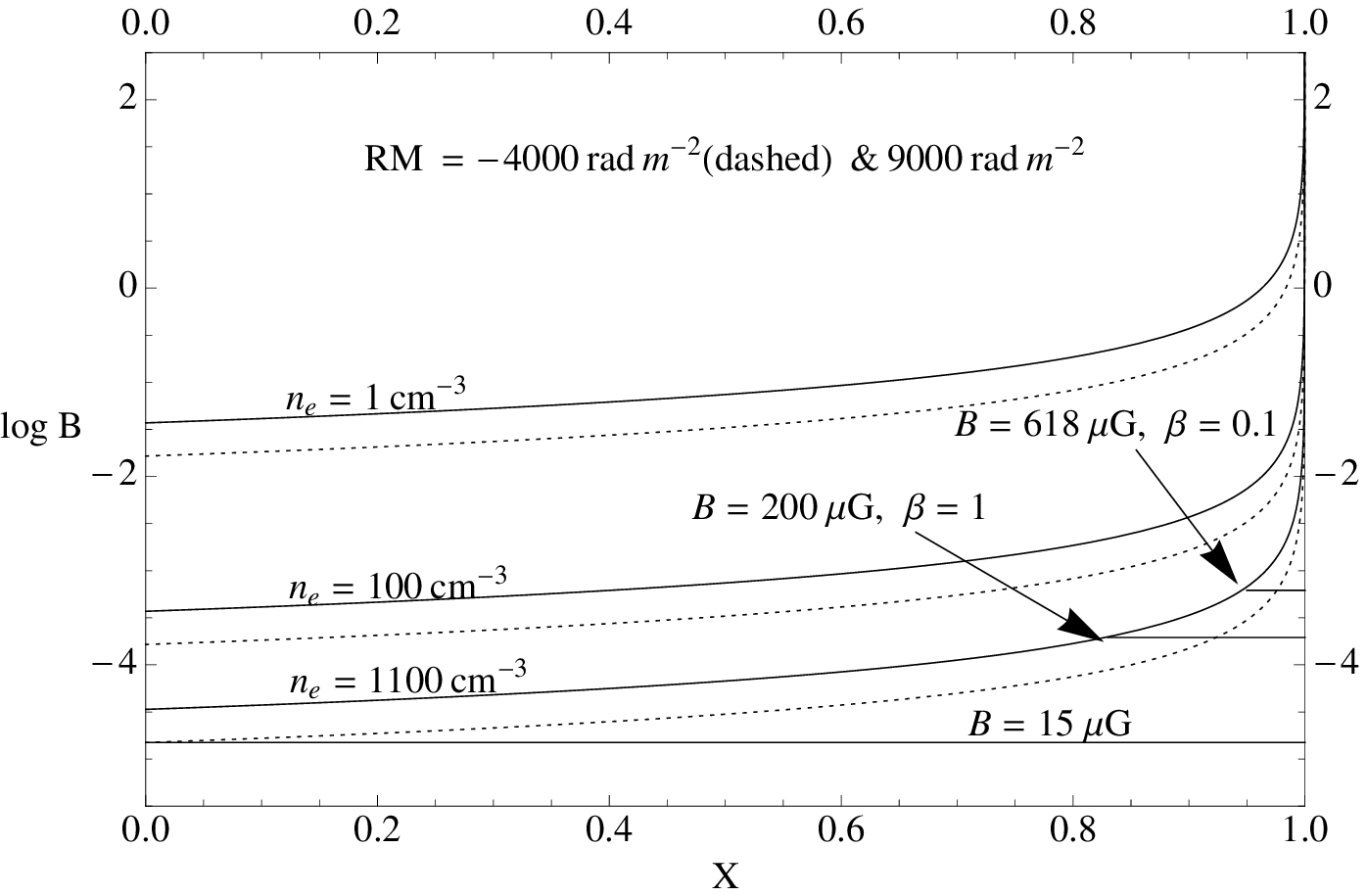}
     \label{fig:log B vs X M87}
   }
   \subfigure[log $B$ vs $X$ for Abell clusters (Eilek \& Owen 2002)]
   {
     \includegraphics[width=8.8cm]{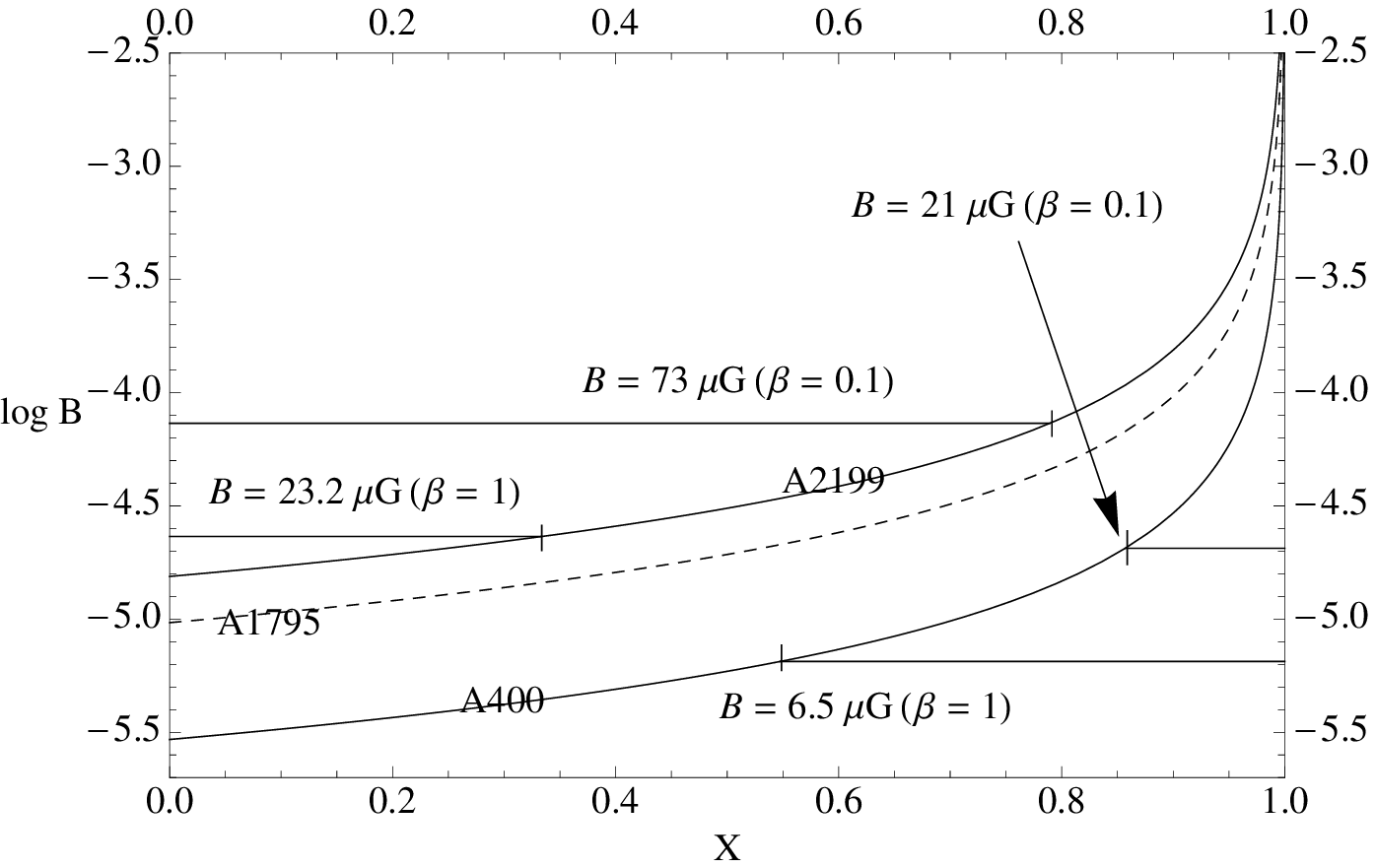}
     \label{fig:ln B vs X (A)}
   }

   }

\caption{These $Log B$ vs $X$ plots were made using RM, $B$,  $n_{e}$ and $T$ from Zavala \& Taylor (2002) and Eilek \& Owen (2002).  The rotation angle for (a) was converted from the RM -4000 rad $m^{-2}$ (dashed line) and 9000 rad $m^{-2}$ (solid line) at $\lambda$= 1.35 $cm$ (22.2 GHz).  For (b), each data point for
 RM (50, 1500, 750 rad $m^{2}$); $n_{e}$
(0.0021, 0.064, 0.02 $cm^{-3}$); and $T$ (1.5, 5.1, 1.2 $keV$)
where the values in parentheses are for A400, A1795, A2199 respectively,
were also plotted for 22.2GHz.  The values of $\beta$ correspond to the particular cluster which the spearate lines of $B$ intersect, explaining why different values of the field correspond to
the same $\beta =0.1$ for A400 and A2199.}
\end{figure*}

If  the RM and two of the three quantities   $n_{e}$, $X$ and $B$ are independently  known, the FR equation is exactly determined.  However, even  if only one variable and RM are known, the other variables can be constrained.

There have been efforts to  interpret the RM  with a subset of independently measured variables
$n_{e}$, $B$ and $T$.  For example, \cite{z3}  calculated the $B$ field in M87 using the RM and
an independently determined $n_{e}$.
For RM = -4000 rad $m^{-2}$, Eq.(\ref{1st order expanded FR}) was used to get
$B\sim$15 $\mu G$, but   $X=0$ was assumed.
We can  revisit the interpretation of the measured RM without assuming $X=0$ a priori.
As seen in Fig.\ref{fig:log B vs X M87},
the cross point of $B\sim$15 $\mu G$ and RM = -4000 rad $m^{-2}$ with $n_{e}$ = 1100$cm^{-3}$ (dashed line) is $X=0$.
In contrast, if we suppose that  $B$ = 200 $\mu G$, which corresponds to the thermal
equipartition condition when $T\sim 10^{4}K$ ($n_{e}k_{B}T=B^{2}/8 \pi$ where $B$ is the mean line of sight field that causes FR, e.g. \cite{g0}), different $X$ values result:  For RM = 9000 rad $m^{-2}$ and -4000 rad $m^{-2}$,
the $X$ values  are 0.83 and 0.92 respectively.  The corresponding $n_{e^{+}}$ values are 913 and 1012 $cm^{-3}$. If instead magnetic pressure dominates (e.g.
dominant ($\beta\equiv P_{g}/P_{B}=8\pi n_{e}k_{B}T/B^2<1$),
then $X$ increases as shown in Fig.3.

As another set of examples showing the degeneracy between $B$ and $X$, we  consider the independently measured  RM, $n_{e}$, $B$ and $T$ for clusters A400, A1795 and A2199 \citep{e1} which respectively include radio sources 3C75, 4C46.42 and 3C338 to obtain Fig.\ref{fig:ln B vs X (A)}.  $B$ is also the mean line of sight field but we note that the definition of the magnetic pressure ($P_{B} = 3<B_{\|}>^{2}/ 8 \pi$) is different from that of \cite{z3}.  We have taken this into account when interpreting their respective data.  There, instead of assuming $X=0$, as is normally done to obtain the $B$, we chose selected field strengths (straight lines on the plot) and identify the constraints this places on $X$ by where these lines intersect with the curves.
For example, for A400,  if magnetic and thermal pressures were in equipartition  ($\beta=1$), the data would imply  $X=0.55$ (Here $n_{e}=0.0021$ $cm^{-3}$, $n_{e^{+}}=1.155\times10^{-3}$ $cm^{-3}$, $n_{i}=9.45\times$ $10^{-4}$ $cm^{-3}$)). However, if  magnetic pressure were dominant (e.g. $\beta=0.1$),
then  $X=0.86$ ($n_{e}=0.0021$ $cm^{-3}$, $n_{e^{+}}=1.806\times10^{-3}$ $cm^{-3}$, $n_{i}=2.94\times$ $10^{-4}$ $cm^{-3}$).  The extent to which $X$ differs form zero in clusters
could depend on how close to a radio source core the RM is measured as one would indeed expect $X=0$ far away from radio jets.
As better spatial coverage of both FR and synchrotron constraints become available for jets, the use of plots such as those of Fig.1 and 2 become more powerful.

\section{Conclusions}
Assuming plasma neutrality, we have generalized the calculation of FR for a neutral
plasma containing  electrons, positrons, and ions. Positrons contribute equally and oppositely to electrons under the approximations considered. Thus   as the ratio of positrons to electrons  $X$ increases from 0 to 1, the FR weakens  for a given line of sight magnetic field and density. Correspondingly, for a given RM,
a larger value of $X$ would imply a field strength and or electron density in excess of that inferred for $X=0$.

While $X=0$ is often assumed for astrophysical plasmas, this assumption is not
necessarily valid for  jets and magnetospheres of black holes and neutron stars, where in fact
it is of interest to independently determine $X$.
We have shown quantitatively the degeneracies in the space of parameters $n_e$, $X$, and $B$
and how they can be constrained.  In principle, if independent data on $B$ and $n_e$ can be obtained, then a given RM measurement can be used to obtain $X$. The pursuit of
$X$ has been particularly elusive in the context of AGN jets and we hope that the
calculations herein provide a tool toward this goal, and help  motivate the pursuit of
further data.

\section*{Acknowledgments} KP acknowledges a Horton Graduate Fellowship
from the Laboratory for Laser Energetics at the Univ. of Rochester.
EGB acknowledges support from
NSF grants AST-0406799
AST-0807363, and NASA grant ATP04-0000-0016.


\begin{thebibliography}{199}
\bibitem[\protect\citeauthoryear{Asada et al.}{2002}]{a1} Asada K., Inoue M., Uchida Y., Kameno S., Fujisawa K., Iguchi S., Mutoh M., 2002, PASJ, 54, L39
\bibitem[\protect\citeauthoryear{Asada et al.}{2008}]{a2} Asada K., Inoue M., Kameno S., Nagai H., 2008, ApJ, 675, 79
\bibitem[\protect\citeauthoryear{Ashok}{2004}]{a3} Ashok D., 2004, Lectures on Electromagnetism, 1st edn., 389p, Rinton Press
\bibitem[\protect\citeauthoryear{Attridge et al.}{2005}]{a4}Attridge J. M., Wardle J. F. C., Homan D. C. 2005, ApJ, 633, L85
\bibitem[\protect\citeauthoryear{Begelman et al.}{1984}]{b1}Begelman M. C.,  Blandford R. D., Rees M. J.,
 1984, Rev. Mod. Phys., 56, 255
\bibitem[\protect\citeauthoryear{Blandford \& Znajek}{1977}]{bz77} Blandford R.~D., Znajek R.~L., 1977, MNRAS, 179, 433
\bibitem[\protect\citeauthoryear{Celotti \& Fabian}{1993}]{c1}
Celotti A., Fabian A.C., 1993, MNRAS, 264, 228
\bibitem[\protect\citeauthoryear{Li et al.}{2006}]{li06} Li
H., Lapenta G., Finn J.~M., Li S., Colgate S.~A., 2006, ApJ, 643, 92
\bibitem[\protect\citeauthoryear{Eilek \& Owen}{2002}]{e1}
 Eilek Jean A., Owen Frazer N., 2002, ApJ, 567, 202
\bibitem[\protect\citeauthoryear{Gabuzda et al.}{2008}]{g0}
Gabuzda D. C., Pushkarev A. B., Garnich N. N., 2001, MNRAS, 327, 1
\bibitem[\protect\citeauthoryear{Gurnett \& Bhattacharjee}{2005}]{g1}
Gurnett D. A., Bhattacharjee A., 2005, Introduction to Plasma Physics, 94p, Cambridge University Press
\bibitem[\protect\citeauthoryear{Ghisellini}{2008}]{g08}
Ghisellini G., 2008, IJMPD, 17, 1491
\bibitem[\protect\citeauthoryear{Hall \& Shukla}{2005}]{h1}
Hall J.O., Shukla P.K., 2005, Phys. Plasma, 12, 084507
\bibitem[\protect\citeauthoryear{Hubbard \& Blackman} {2006}]{hubbard}
Hubbard, A., \& Blackman, E.G.,  2006, MNRAS, 371, 1717
\bibitem[\protect\citeauthoryear{Jaffe}{1980}]{j1}
Jaffe W., 1980, ApJ. 241, 925
\bibitem[\protect\citeauthoryear{Kulsrud}{2005}]{k1}
Kulsrud R. M., 2005, Plasma Physics for Astrophysics, 262p, Princeton University Press
\bibitem[\protect\citeauthoryear{Kronberg, Perry,
\& Zukowski}{1990}]{kpz90} Kronberg P.~P., Perry J.~J., Zukowski E.~L.~H., 1990, ApJ, 355, L31
\bibitem[\protect\citeauthoryear{Livio, Ogilvie \& Pringle}{1999}]{lop99} Livio M., Ogilvie G.~I., Pringle J.~E., 1999, ApJ, 512, 100
\bibitem[\protect\citeauthoryear{Michel}{1982}]{m2}Michel F. C., 1982, Rev. Mod. Phys., 54, 1
Manchester R. N., Taylor J. H., 1978, Pulsars, 1st edn., 134p, W.H.Freeman \& Co Ltd
\bibitem[\protect\citeauthoryear{Reynolds et al.}{1996}]{r1}
Reynolds, C.S., Fabian, A.C., Celotti, A., Rees, M.J. 1996b, MNRAS, 283, 873
\bibitem[\protect\citeauthoryear{Savolainen et al.}{2006}]{t1}
    Savolainen T., Wiik K., Valtaoja E., Tornikoski M., 2006, ASP, 446, 71-85
\bibitem[\protect\citeauthoryear{Skilling}{1971}]{s71}
Skilling J., 1971, PhFl, 14, 2523
\bibitem[\protect\citeauthoryear{Zavala \& Taylor}{2005}]{z1}
    Zavala R. T., Taylor G. B., 2005, ApJ, 626, L73
\bibitem[\protect\citeauthoryear{Zavala \& Taylor}{2001}]{z2}
    Zavala R. T., Taylor G. B., 2001, ApJ, 550, L147
\bibitem[\protect\citeauthoryear{Zavala \& Taylor}{2002}]{z3}
    Zavala R. T., Taylor G. B., 2002, ApJ, 566, L9-L12

\end{thebibliography}
\end{document}